# Promise and challenges of heart chamber segmentation from non-contrast CT scans using contrastive unpaired image translation: a feasibility study

Jing Wang[1], Tong Yu[1,2], Hao-En Lu[1], Zixue Zeng [1, 2], Joseph K. Leader [1], Xin Meng[1], Jianbing Zhu[1], Jiantao Pu[1, 2, 3]

[1] Department of Radiology, University of Pittsburgh, Pittsburgh, PA 15213, USA

[2] Department of Bioengineering, University of Pittsburgh, Pittsburgh, PA 15213, USA

[3] Department of Ophthalmology, University of Pittsburgh, Pittsburgh, PA 15213, USA

*Corresponding authors and guarantors of the entire manuscript:

Jiantao Pu, PhD

Contact Phone: (412) 641-2571

E-mail address: jip13@pitt.edu

Mailing address: 3240 Craft Place, Pittsburgh, PA  15213

## ABSTRACT

**Purpose:** To evaluate the feasibility and challenges of heart chamber segmentation from non-contrast CT scans using contrastive unpaired image translation and deep learning-based segmentation.

**Approach:** We developed ChameleonNet, a framework utilizing the Contrastive Unpaired Translation (CUT) network with decoupled contrastive learning (DCL) loss to synthesize non-contrast CT from contrast CT scans. Using annotations of four heart chambers (left atrium (LA), left ventricle (LV), right atrium (RA), and right ventricle (RV)) from contrast scans, we trained a Hausdorff distance loss-enhanced nnU-Net on synthesized non-contrast images. The translation model was trained with 35,538 contrast-enhanced and 37,197 non-contrast CT slices. The segmentation model was trained with 292 synthesized non-contrast scans. Performance was evaluated using Dice similarity coefficient (DSC) and 95th Hausdorff distance (HD95) on 36 synthesized non-contrast scans, and volume agreement on 36 real non-contrast CT scans was assessed using Pearson correlation, mean absolute percentage error (MAPE), and mean percentage error (MPE).

**Results:** The segmentation model achieved DSC of 0.94 (±0.01), 0.91 (±0.04), 0.92 (±0.03), 0.93 (±0.02), and HD95 of 3.63 (±1.49), 5.74 (±4.08), 5.18 (±1.77), 5.51 (±3.21) mm on synthesized non-contrast images for LA, LV, RA, and RV, respectively. On real non-contrast CT scans, Pearson correlations were 0.93, 0.82, 0.87, and 0.89 (all $p<0.001$), with MAPE ranging from 9.22% to 20.79%, and MPE ranging from -12.52% to 4.67%.

**Conclusions:** ChameleonNet demonstrated feasibility for heart chamber segmentation from non-contrast CT without manual non-contrast annotations. However, volume errors, particularly for LV and RV, indicate that further refinement and validation are needed before clinical use.

**Keywords:** Computed tomography, Heart, Image translation, Image segmentation

# 1. INTRODUCTION

Cardiovascular diseases are a leading cause of mortality and morbidity worldwide[1]. Heart chamber sizes are closely associated with various cardiovascular conditions and can predict the risk of events such as ischemic heart disease[2], atrial fibrillation[3], and heart failure[4].

Cardiac imaging is essential for assessing cardiac function, visualizing heart structures, and supporting diagnosis and management. Contrast-enhanced CT scans provide clear anatomical details through contrast agents, whereas non-contrast CT scans lack sufficient tissue differentiation for accurate heart chamber identification, making manual annotation highly challenging. However, accurate identification of cardiac chambers from non-contrast CT scans remains highly desirable for several reasons. First, non-contrast CT offers reduced costs, greater accessibility, and a lower risk of complications compared to contrast-enhanced CT. Second, non-contrast CT scans are widely utilized in routine clinical practice for the diagnosis and screening of many other diseases, such as lung cancer, chronic obstructive pulmonary disease (COPD), and asthma, providing opportunistic screening of heart disease without the need for additional patient enrollment or imaging procedures. Moreover, accurate segmentation of heart substructures is valuable in optimizing radiation therapy[5], as dose to heart substructures is linked to adverse outcomes including non-cancer death[6].

However, manual delineation of heart chambers on non-contrast CT scans is considerably more challenging than on contrast-enhanced CT scans, due to the poor contrast between adjacent soft tissues, which may increase interobserver variability. To overcome this issue, computerized algorithms have been developed to automate the segmentation process and reduce inter-observer variability. Multi-atlas-based methods, which segment heart substructures by aligning the target image with one or more atlas samples, are widely used[7,8]. However, the segmentation results are often compromised by limited atlas generalizability and the effectiveness of registration methods. Deep learning methods, such as nnU-Net[9,10], have demonstrated exceptional performance in medical image segmentation. For instance, Chen et al.[11] trained 3D U-Net and nnU-Net models using ground truth labels of 19 cardiac substructures, which were delineated by two radiation oncologists on non-contrast CT scans. However, the substantial variability between observers raises concerns about the reliability of manual labeling. The scarcity of annotations on non-contrast CT scans has prompted efforts to utilize labels from different imaging modalities. Some studies[12,13] employed annotations from well-aligned contrast CT as ground truth, and some trained models with virtual non-contrast (VNC) images generated from contrast-enhanced scans using dual-energy computed tomography[14]. However, in practice, contrast-enhanced CT scans and non-contrast CT scans acquired on the same subjects are not perfectly paired or aligned, making it difficult to leverage annotations from contrast-enhanced counterparts for developing robust segmentation models. Recently, image translation techniques have been used to address this issue of insufficient labeling in medical image related applications. For instance, Veit Sandfort et al.[15] employed CycleGAN[16] to generate synthesized non-contrast images from contrast CT scans, subsequently training segmentation models with contrast CT augmented with synthesized images. This approach improved segmentation performance for kidney, liver, and spleen on non-contrast datasets. However, most studies addressing domain shifts to compensate for the lack of annotations for heart substructure segmentation focus on MRI applications[17,18] , with minimal focus on transforming contrast CT scans into non-contrast CT scans.

In this study, we introduce ChameleonNet, a two-stage feasibility framework that combines contrastive unpaired image translation with deep-learning-based segmentation to explore heart chamber segmentation from non-contrast CT scans. First, an unsupervised image translation network was employed to generate synthesized non-contrast CT scans from contrast-enhanced CT scans while maintaining the integrity of anatomical structures and thus precise image-mask correspondence. The image translation network employed the contrastive unpaired translation network (CUT)[19] but with a decoupled training loss to improve the efficiency of the pull-push mechanism in contrastive learning. Next, a modified nnU-Net model was trained on the synthesized non-contrast CT scans to segment four heart chambers: the left atrium (LA), left ventricle (LV), right atrium (RA), and right ventricle (RV). By leveraging accurate annotations from contrast-enhanced CT scans and eliminating the need for unreliable labels on non-contrast CT scans, our approach offers a reliable method for estimating cardiac chambers on non-contrast CT scans, with potential applications in the prognosis and diagnosis of cardiovascular diseases. The key features of this work are: (1) the use of a decoupled contrastive learning loss within the CUT framework, which accelerates translation training by eliminating the negative-positive coupling effect; (2) a large-scale training dataset with a fully independent test set of real non-contrast CT scans; and (3) a CT-to-CT domain adaptation approach that preserves the native resolution and HU calibration of CT imaging.

## 2. MATERIALS AND METHODS

### 2.1 Study Cohort

A total of 550 contrast chest CT scans and 550 non-contrast chest CT scans were randomly selected from the datasets in our previous studies[20-23] for the image translation stage, with 50 scans from each group reserved for testing (Fig. 1). The training and test sets were split at the patient level, ensuring that no patient contributed slices to both sets. All 1,100 CT scans were reconstructed to a slice thickness of 2.5 mm, and axial slices were extracted, resulting in 35,538 contrast and 37,197 non-contrast CT slices for training, and 3,419 contrast-enhanced and 3,784 non-contrast CT slices for testing.

In the segmentation stage, the LA, LV, RA and RV were manually delineated on a different set of 325 contrast CT scans by a radiologist with experience in cardiac imaging. Of the 325 annotated scans, 292 were used for training and 33 for validation. An independent set of 36 paired contrast and non-contrast CT scans were collected to evaluate segmentation performance on real non-contrast images. The radiologist's manual annotations on the contrast-enhanced CT scans served as references for evaluating segmentation accuracy.

For testing, the contrast-enhanced CT scans were translated to non-contrast CT scans to assess segmentation performance on the synthesized non-contrast CT scans. Real non-contrast CT scans were used to evaluate the segmentation performance on actual data. Notably, although the paired contrast and non-contrast CT scans were acquired from the same patients, they were not ECG-gated, resulting in misaligned image pairs.

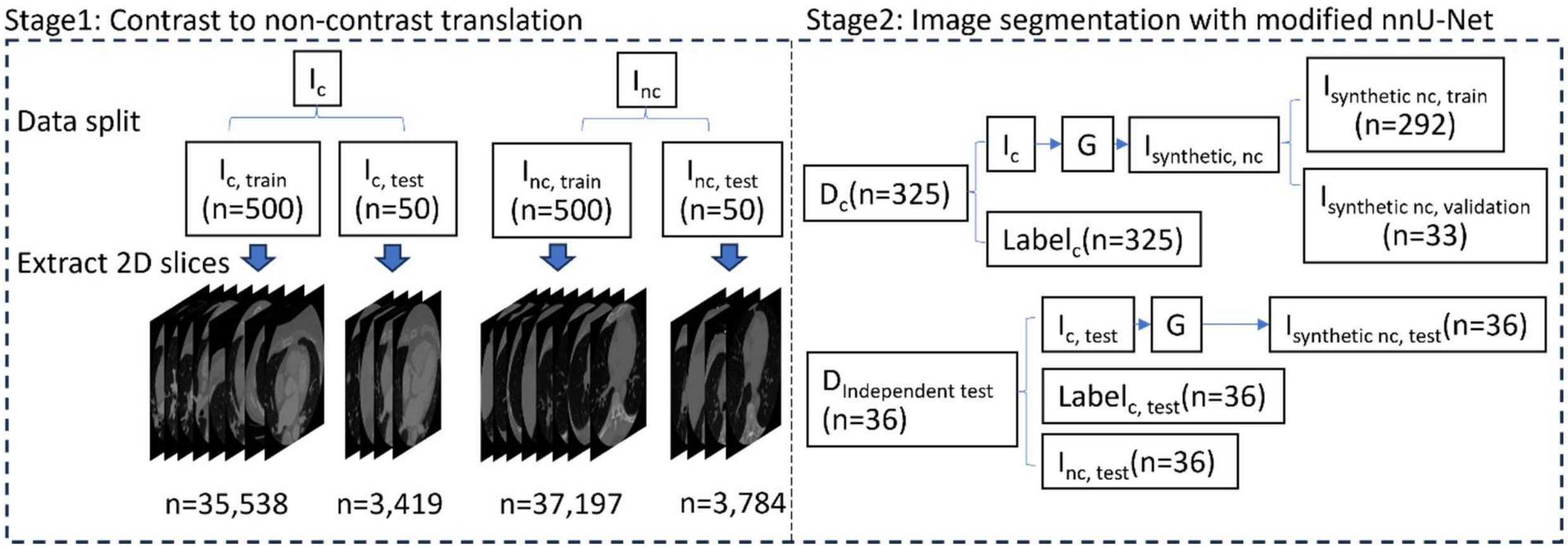

**Figure 1.** Data splitting for model training, evaluation and testing. In Stage 1, contrast-enhanced CT scans $I_c$ and non-contrast CT scans $I_{nc}$ were divided into training and test set, followed by the extraction of 2D axial slices. In Stage 2, scans in contrast-enhanced dataset $D_c$ were translated by the generator $G$ into non-contrast images. The independent test set consists of contrast and non-contrast CTs from the same patient, along with radiologist's annotations on the contrast-enhanced CT scans.

**2.2 ChameleonNet Overview**

The workflow of our proposed ChameleonNet is illustrated in Fig. 2. The method utilizes a two-stage approach, leveraging a generative adversarial network (GAN) and a segmentation network. The GAN comprises a generator $G$ and a discriminator $D$, trained iteratively to achieve image translation.

In the image translation stage, the goal is to translate contrast-enhanced images $I_c$into synthetic non-contrast images $I_{nc}$, while preserving anatomical details. Using unpaired datasets of $I_c$ and $I_{nc}$, the translation network learns the feature distribution of non-contrast images and generates synthetic non-contrast CT images that closely resemble real non-contrast images without compromising the structure integrity of the original contrast-enhanced images. In the image segmentation stage, a modified nnU-Net model was trained to segment LA, LV, RA, and RV, using the synthetic non-contrast image and the annotations of the heart chambers on contrast-enhanced images.

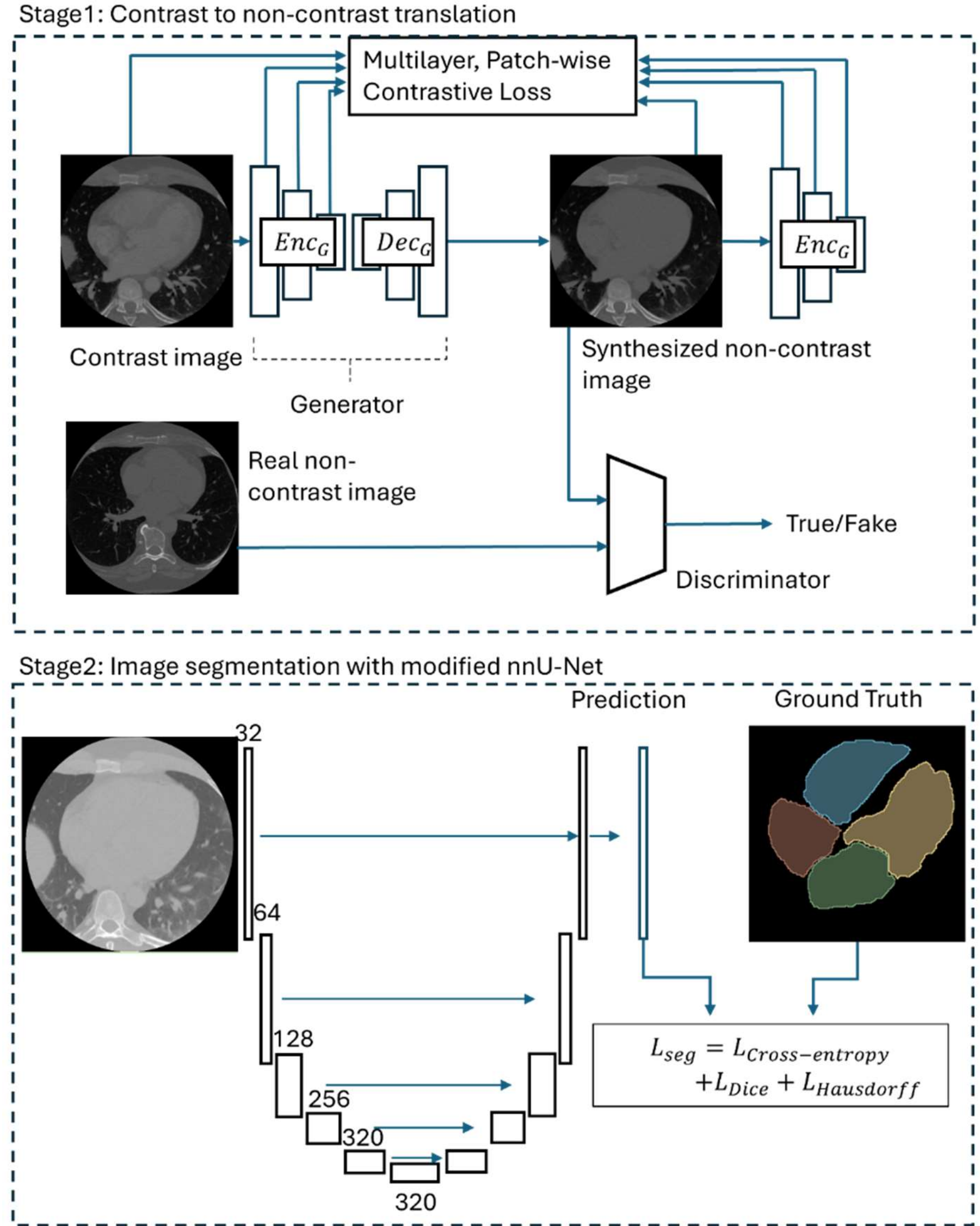

**Figure 2**. Schematic diagram illustrating ChameleonNet. Stage 1: The generator translates a contrast-enhanced image into a synthetic non-contrast image. Multilayer feature maps are extracted, and a patch-wise decoupled contrast learning loss is computed. The discriminator distinguishes between real and synthetic non-contrast images. Stage 2: A modified nnU-Net segments LA, LV, RA, and RV from the synthetic non-contrast image, minimizing Dice, cross-entropy, and an additional Hausdorff distance loss between predictions and the contrast image annotations.

### 2.3 Stage I: Contrast to Non-contrast Translation

Our image translation method built on the CUT architecture[19] but enhanced it with a novel loss function. The generator takes a contrast-enhanced image as input and generates a synthetic non-contrast image. Feature patches from multiple layers of both the input contrast image and the synthesized non-contrast image are extracted, and a multi-layer patch-wise contrastive loss is computed with improved training efficiency by removing the coupling effect between positive and negative pairs in contrastive learning. The discriminator distinguishes between real and synthesized non-contrast images to calculate the adversarial loss.

### 2.3.1 Patch-wise contrastive loss

The generator $G$ consists of an encoder $Enc_G$ and a decoder $Dec_G$, where the synthesized image is $\hat{y} = G(x) = Dec_G(Enc_G(x))$. Following the CUT method[19], an anchor patch in the generated image $\hat{y}$ with feature vector $v$ is matched to its corresponding patch in the input image $x$ at the same location as a positive sample with feature vector $v^+$. Other patches in $x$ are treated as negative samples with feature vector $v^-$. The loss is designed to maximize the mutual information between the anchor and its positive sample while maximizing the distance to the negative samples, facilitating learning the shared features between the source and target domains. For anchor with feature vector $v$, the original patch-wise contrastive loss is formulated as:

$$\ell_{\text{InfoNCE}}(v, v^+, v_n^-) = -\log \frac{\exp(v \cdot v^+/\tau)}{\exp(v \cdot v^+/\tau) + \sum_{n=1}^{N} \exp(v \cdot v_n^-/\tau)} \tag{1}$$

where $\tau = 0.07$ is a scaling factor, following the default value used in the original CUT paper[19].

### 2.3.2 Decoupled contrastive learning for training efficiency

The negative-positive coupling (NPC) effect, observed in InfoNCE loss[19,24], leads to reduced learning efficiency, especially when negative samples are too distant from the anchor or the positive pair is too close[25]. The gradients of the $\ell_{\text{InfoNCE}}$ are expressed as:

$$\begin{cases} -\nabla_{v}\ell_{\text{InfoNCE}} = \dfrac{q_{NPC}}{\tau}\left[v^+ - \dfrac{\sum_{n=1}^{N} v_n^- \cdot \exp(v \cdot v_n^-/\tau)}{\sum_{n=1}^{N} \exp(v \cdot v_n^-/\tau)}\right] \\ -\nabla_{v^+}\ell_{\text{InfoNCE}} = \dfrac{q_{NPC}}{\tau} \cdot v \\ -\nabla_{v_n^-}\ell_{\text{InfoNCE}} = -\dfrac{q_{NPC}}{\tau} \cdot \dfrac{\exp(v \cdot v_n^-/\tau)}{\sum_{k=1}^{N} \exp(v \cdot v_k^-/\tau)} \end{cases} \tag{2}$$

where

$$q_{NPC} = 1 - \frac{\exp(v \cdot v^+/\tau)}{\exp(v \cdot v^+/\tau) + \sum_{n=1}^{N} \exp(v \cdot v_n^-/\tau)} \tag{3}$$

This multiplier is less than 1 and it approaches zero when the similarity between anchor and the positive sample is high or when the negative pairs are easily distinguishable, leading to vanishing gradients during model training. To address this, a decoupled contrastive learning loss is introduced by removing the positive pair term from the denominator of $\ell_{\text{InfoNCE}}$[25]. With this adjustment, the NPC multiplier $q_{NPC} = 1$ when training with $\ell_{\text{DCE}}$.

$$\ell_{\text{DCE}}(v, v^{+}, v_{n}^{-}) = -\log \frac{\exp(v \cdot v^{+}/\tau)}{\sum_{n=1}^{N} \exp(v \cdot v_{n}^{-}/\tau)} \tag{4}$$

### 2.3.3 Multilayer sampling

Patches from multiple layers of the encoder in the generator $Enc_G$ were sampled and used in contrastive learning to capture patch similarity across different scales[19]. Both the input image $x$ and its generated counterpart in the target domain, $\hat{y}$, were passed to the generator. Features were extracted from $L$ layers of interest and processed through a two-layer multi-layer perceptron (MLP) network. Denoting $v_{l}^{n}$ for feature of the $n$th location at layer $l$, $v_{l}^{N\backslash n}$ for features at other locations, the multilayer features for $x$ are represented with $\{v_l\}_L$ and the multilayer feature for $\hat{y}$ is $\{\hat{v}_l\}_L$, where $l \in \{1,2,\dots,L\}$. The multilayer decoupled contrastive learning loss is then formulated as:

$$\mathcal{L}_{\text{DCE}}(X) = \mathbb{E}_{x\sim X} \sum_{l=1}^{L} \sum_{n=1}^{N} \ell_{\text{DCE}}(\hat{v}_{l}^{n}, v_{l}^{n}, v_{l}^{N\backslash n}) \tag{5}$$

An identity loss is also incorporated for images in domain $Y$ to prevent unnecessary alterations introduced by the translation network:

$$\mathcal{L}_{\text{DCE}}(Y) = \mathbb{E}_{y\sim Y} \sum_{l=1}^{L} \sum_{n=1}^{N} \ell_{\text{DCE}}(\hat{v}_{l}^{n}, v_{l}^{n}, v_{l}^{N\backslash n}) \tag{6}$$

### 2.3.4 Full loss

With consideration of the adversarial loss of GAN[26], the full loss for the image translation model is:

$$\mathcal{L}_{\text{Total}} = \mathcal{L}_{\text{GAN}} + \lambda_X \mathcal{L}_{\text{DCE}}(X) + \lambda_Y \mathcal{L}_{\text{DCE}}(Y) \tag{7}$$

where:

$$\mathcal{L}_{\text{GAN}} = \mathbb{E}_{y\sim Y} \log D(y) + \mathbb{E}_{x\sim X} \log(1 - D(G(x))) \tag{8}$$

### 2.4 Stage II: Segmentation with Hausdorff Distance Loss Modified nnU-Net

The image segmentation model was trained with synthesized non-contrast CT images and the ground truth segmentation from the corresponding contrast-enhanced CT scans with the nnU-Net architecture (Fig. 2). Two model variants were trained for comparison. The standard nnU-Net used the default combined cross-entropy and Dice loss function. The modified nnU-Net additionally incorporated a Hausdorff distance loss term, such that model predictions were compared to ground truth using cross-entropy loss, Dice loss, and Hausdorff distance loss.

Given two point sets, P and Q, the one-sided HD from P to Q and from Q to P are defined as:

$$hd(P,Q) = \max_{p \in P} \min_{q \in Q} \|p - q\|_2 \tag{9}$$

$$hd(Q,P) = \max_{q \in Q} \min_{p \in P} \|q - p\|_2 \tag{10}$$

The Hausdorff distance loss is defined as:

$$HD(P,Q) = \max\big(hd(P,Q), hd(Q,P)\big) \tag{11}$$

The Hausdorff distance calculation was approximated using distance transformation[27] in the implementation.

The loss for the segmentation model is:

$$L_{\text{Seg}} = L_{\text{Cross-Entropy}} + L_{\text{Dice}} + \lambda_{HD} L_{\text{Hausdorff}} \tag{12}$$

where $L_{\text{Cross-Entropy}}$, $L_{\text{Dice}}$ and $L_{\text{Hausdorff}}$ represent cross-entropy loss, Dice loss and Hausdorff distance loss respectively. $\lambda_{HD}$ is the weight for the Hausdorff distance loss.

### 2.5 Model Training

In the image translation stage, we replaced the InfoNCE loss in CUT[19] with DCE loss. The generator was based on a ResNet with 9 residual blocks, and training used a PatchGAN discriminator[28] with Least Square GAN loss[29]. For input images of size 512x512, we sampled 1024 patches per feature layer, with other hyperparameters remaining the same as those in[19] ($\lambda_X = \lambda_Y = 1$). The model was trained using the Adam optimizer with an initial learning rate at 0.0002, remaining constant for the first 5 epochs and linearly decaying to 0 over the next 5 epochs. CUT and CycleGAN were trained with default settings[16,19], except for the 1024 patch sampling in CUT. For segmentation, the nnU-Net model was trained with random 90*160*60 image crops, using pixel spacing [2.5, 0.49, 0.49]. The 3D U-Net has a depth of 6 and feature sizes [32, 64, 128, 256, 320, 320]. Training was performed with the stochastic gradient descent (SGD) optimizer, starting at a learning rate of 0.01 and decaying to 0 over 250 epochs using a polynomial function. Hausdorff distance loss was integrated into the deep supervision process[30], with a weight of 0.02. This value was selected empirically by evaluating weights of 0.01, 0.02, and 0.05 on the validation set, with $\lambda_{HD} = 0.02$ yielding the best segmentation performance. All the models were trained on NVIDIA GeForce RTX 3090.

### 2.6 Performance Evaluation

#### 2.6.1 Contrast to non-contrast image translation

The performance of image translation in the first stage was compared with two baseline models: CycleGAN[16] and CUT[19], using Structural Similarity Index Measure (SSIM)[31] and Fréchet inception distance (FID)[32] metrics. SSIM was computed by comparing each synthesized non-contrast image against its source contrast-enhanced image to assess structural preservation during translation, while FID was computed by comparing the feature distribution of synthesized non-contrast images against the distribution of real non-contrast images to evaluate realism and domain alignment. Together, high SSIM and low FID provide evidence that the translation network both preserves anatomical structure and produces images whose intensity distribution closely matches real non-contrast CT.

SSIM evaluates image similarity by quantifying the degradation of structural information. It compares image $x$ and $y$ based on three factors: luminance $l(x, y)$, contrast $c(x, y)$, and structure $s(x, y)$ to mimic the human vision system (HVS). The SSIM is defined as:

$$SSIM(x, y) = l(x, y)c(x, y)s(x, y) \tag{13}$$

where

$$\begin{cases} l(x, y) = \dfrac{2\mu_x\mu_y + C_1}{\mu_x^2 + \mu_y^2 + C_1} \\ c(x, y) = \dfrac{2\sigma_x\sigma_y + C_2}{\sigma_x^2 + \sigma_y^2 + C_2} \\ s(x, y) = \dfrac{\sigma_{xy} + C_3}{\sigma_x\sigma_y + C_3} \end{cases} \tag{14}$$

$\mu$ and $\sigma^2$ represent the mean and variance of pixel intensities in an image, while $\sigma_{xy}$ is the covariance between images $x$ and $y$. Constants $C_1$, $C_2$, and $C_3$ prevent division by zero. SSIM value ranges from 0 to 1, with 1 indicating identical images. SSIM was computed for each image slice before and after image translation in the test set. A two-sided paired t-test was conducted to assess significance, with $p<0.05$ considered statistically significant.

FID measures image generation quality by comparing feature distributions between generated and real target images. Assuming two multivariate Gaussian distributions $X_1 \sim N(\mu_1, c_1)$ and $X_2 \sim N(\mu_2, c_2)$, the FID score is calculated as:

$$\text{FID score} = \|\mu_1 - \mu_2\|^2 + \text{Tr}\left(c_1 + c_2 - 2\sqrt{c_1 c_2}\right) \quad (15)$$

where $X_1$ and $X_2$ are the activations extracted from a pretrained Inception model. $\mu_1$ and $\mu_2$ are the feature-wise means of the real and generated images. $c_1$ and $c_2$ are their covariance matrices. Tr denotes the trace of a matrix. Lower FID values indicate better similarity between generated and real non-contrast images.

Contrast-enhanced images and their synthesized non-contrast counterparts were displayed for qualitative evaluation, with a window level of 50 HU and a window size of 500 HU. A difference map was generated by subtracting the input image intensity from the output image, with red indicating increased intensity and blue indicating decreased intensity.

**2.6.2 Image segmentation with enhanced nnU-Net**

Segmentation performance was evaluated with the Dice similarity coefficient (DSC) that measures the agreement between the predicted segmentation and ground truth, and the 95th Hausdorff distance (HD95) that measures surface distance. For two segmentations $A$ and $B$, the DSC is defined as:

$$DSC(A, B) = \frac{2|A \cap B|}{|A| + |B|} \quad (16)$$

where $|A|$ and $|B|$ are the number of pixels in $A$ and $B$.

The HD95 is calculated by finding the 95th percentile of the maximum distance between two point sets $P$ and $Q$ :

$$HD95(P, Q) = \max\left(P_{95}(hd(P, Q)), P_{95}hd(Q, P)\right) \quad (17)$$

where $P_{95}$ refers to the 95th percentile.

The segmentation model was applied to real non-contrast CT scans. Volume difference between the segmentation results and the annotations from contrast CT scan of the same patient were calculated. Agreement was assessed by Pearson correlation, with p-value less than 0.05 indicating statistical significance. Additionally, the mean absolute percentage error (MAPE) and the mean percentage error (MPE) were used to assess segmentation accuracy.

## 3. RESULTS

### 3.1 Contrast to non-contrast image translation

The image generation quality metrics of our method and its comparison with baseline methods are summarized in Table 1 and Fig 3. Our method achieved a mean SSIM of 0.96 (±0.03) and an FID of 13.29, outperforming the baseline models consistently in both metrics. Paired two-sided t-test demonstrated statistical significance ($p<0.05$) when comparing our method with the baselines. The one-sided contrastive image translation structure, including both CUT method and ours, reduced training time by approximately 36% and 39% compared to CycleGAN. Additionally, improved learning efficiency was observed in the FID curves across the three methods over 10 epochs (Fig. 3 (B)).

**Table 1**. Comparison of image translation performance across three models: our method, CycleGAN[16], and CUT[19]. SSIM and FID metrics were evaluated with the test set at epoch 10.

| Model | Training time (hrs/epoch) | SSIM ↑, Mean (Std) | FID ↓ |
|---|---|---|---|
| CycleGAN | ≈ 6.6 | 0.93(0.03) | 14.22 |
| CUT (NCE) | ≈ 4.2 | 0.95(0.03) | 16.21 |
| Ours (DCE) | ≈ 4 | 0.96(0.03) *+ | 13.29 |

**Note:** * indicates statistical significance compared to CycleGAN ($p<0.05$). + indicates statistical significance compared to CUT(NCE) ($p<0.05$).

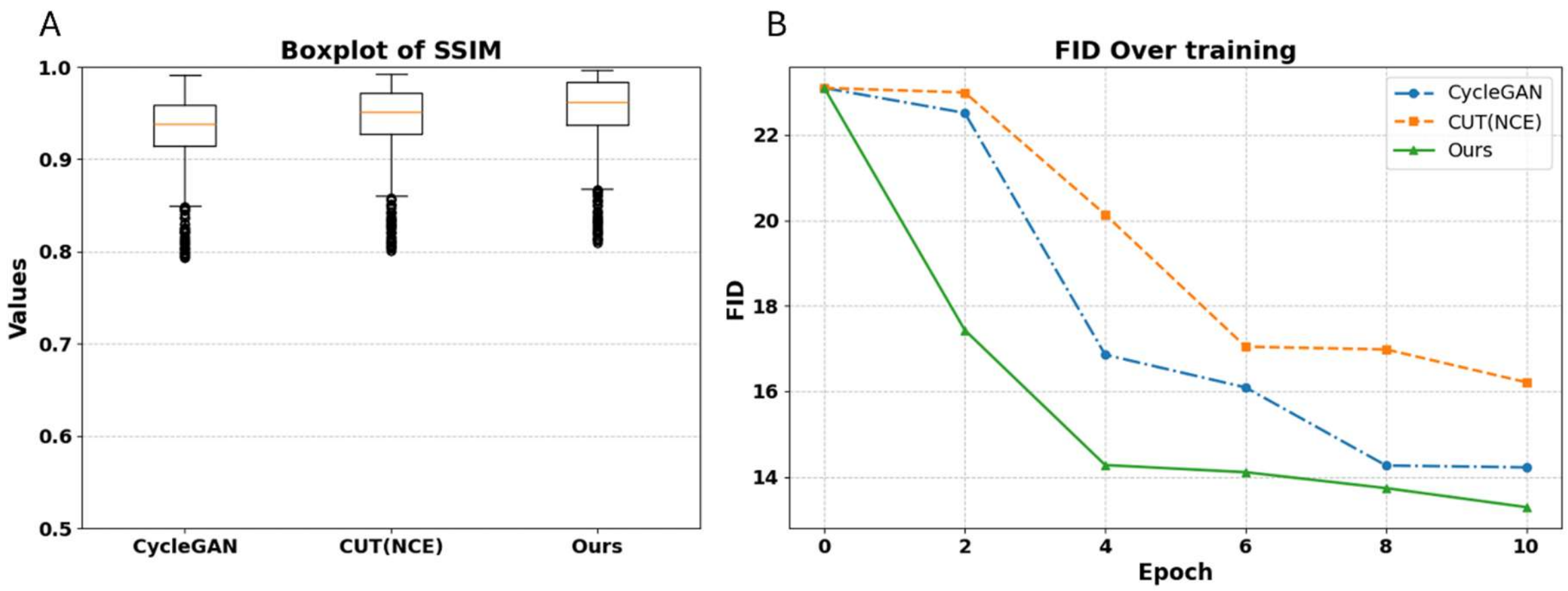

**Figure 3**. (A): SSIM metrics on the test set. (B): FID values over the training process.

Fig. 4 shows two examples of contrast-to-non-contrast image translation, providing a qualitative comparison of image translation performance. All three methods effectively reduced image intensities in contrast-enhanced regions. However, differences in translation quality were observed in the difference maps. CycleGAN introduced intensity alterations along structural edges, with red or blue accumulations at the heart-lung boundary. Meanwhile, non-contrast images generated by CUT increased intensities in the lung area, highlighted in red on the difference map.

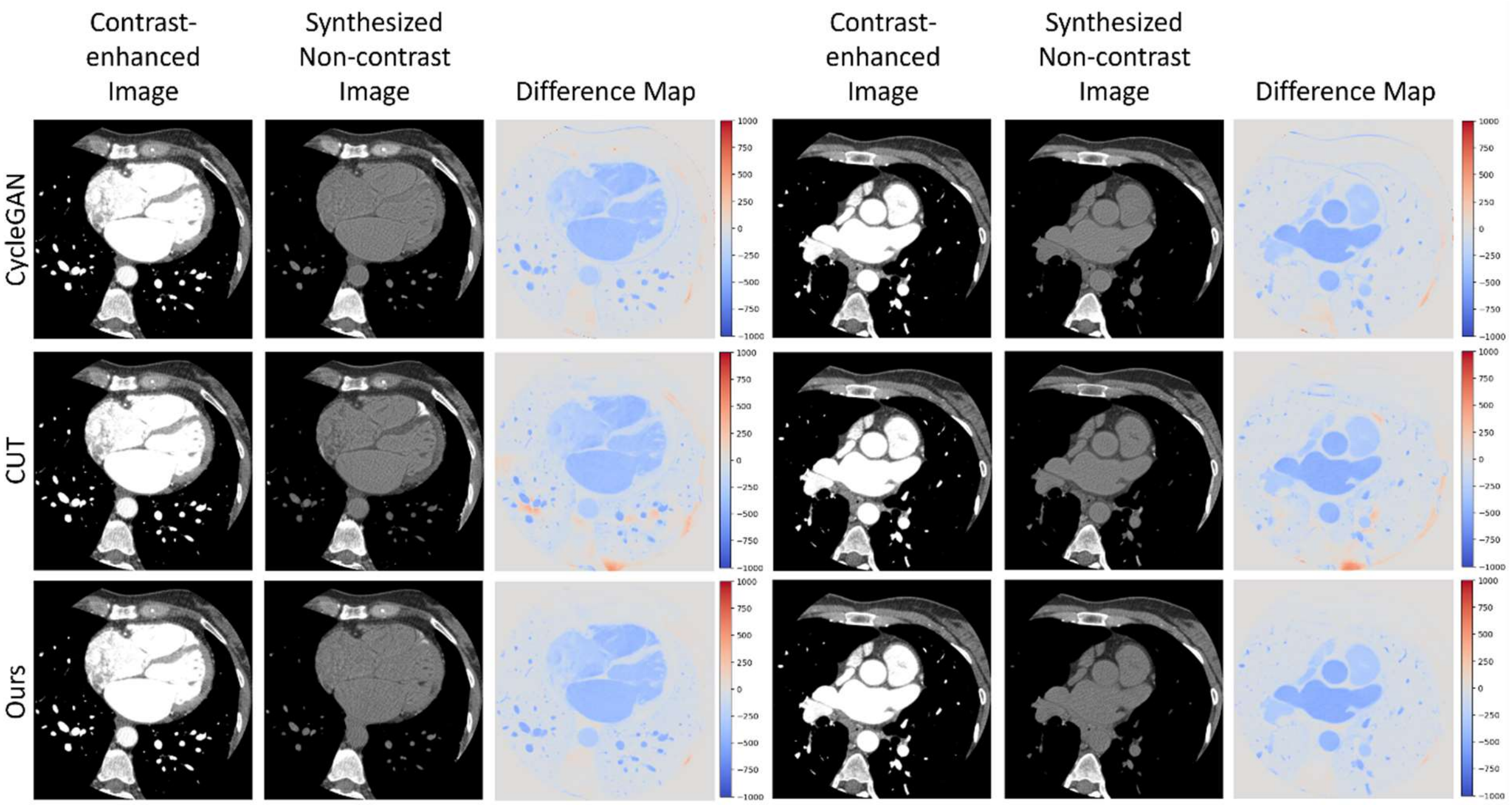

**Figure 4**. Examples of contrast-enhanced to non-contrast image translation. Columns from left to right: original contrast-enhanced images, translated non-contrast images, and corresponding difference maps.

### 3.2 Image segmentation performance

#### 3.2.1 Effect of Hausdorff Distance Loss

Table 2 summarizes segmentation performance on synthetic non-contrast images. Using the same synthetic images generated by the CUT method with DCE, the standard nnU-Net model (SegM1) and the modified nnU-Net model with Hausdorff distance loss (SegM2) achieved comparable DSC and HD95 values. SegM2 achieved mean DSC values of 0.94 (±0.01), 0.91 (±0.04), 0.92 (±0.03), and 0.93 (±0.02) for LA, LV, RA and RV, respectively, with corresponding HD95 values of 3.63 (±1.49), 5.74 (±4.08), 5.18 (±1.77), and 5.51 (±3.21) mm. Differences between SegM1 and SegM2 were small on synthetic images, with statistical significance observed only for RV DSC ($p < 0.05$).

On real non-contrast CT scans, however, the benefit of Hausdorff distance loss was more evident (Table 3, Fig. 5). SegM2 substantially improved LV volume agreement, increasing the Pearson correlation from 0.73 to 0.82 and reducing MAPE from 34.75% to 20.79%. For LA, SegM2 modestly reduced MAPE from 10.07% to 9.22% and MPE from 6.52% to 4.67%, suggesting lower systematic bias. RA and RV performance was similar between the two models. Overall, Hausdorff distance loss provided limited improvement on synthetic test images but improved volume agreement for selected chambers, particularly the LV, on real non-contrast CT scans.

**Table 2.** Dice similarity coefficient and 95th Hausdorff distance metrics for segmentation performance on the synthesized non-contrast test set (n=36).

| Method | CUT(DCE)+SegM1 | | Ours (CUT(DCE)+SegM2) | | CycleGAN+SegM2 | | CUT(NCE)+SegM2 | |
|---|---|---|---|---|---|---|---|---|
| Cardiac Chamber | DSC | HD95 (mm) | DSC | HD95 (mm) | DSC | HD95 (mm) | DSC | HD95 (mm) |
| Left Atrium | 0.94 (0.01) | 3.69 (1.85) | 0.94 (0.01) | 3.63 (1.49) | 0.95 (0.02) | 3.49 (1.45) | 0.95 (0.02) * | 3.37 (2.07) * |
| Left Ventricle | 0.91 (0.06) | 6.20 (3.00) | 0.91 (0.04) | 5.74 (4.08) | 0.89 (0.06) * | 6.33 (2.77) * | 0.91 (0.02) * | 5.09 (3.39) |
| Right Atrium | 0.92 (0.03) | 5.36 (2.47) | 0.92 (0.03) | 5.18 (1.77) | 0.91 (0.03) * | 5.70 (2.18) * | 0.92 (0.03) | 4.96 (1.62) |
| Right Ventricle | 0.93 (0.02) * | 6.13 (4.52) | 0.93 (0.02) | 5.51 (3.21) | 0.91 (0.03) * | 6.92 (5.08) * | 0.92 (0.03) | 6.49 (5.64) |

**Note:** Performance was presented as mean (standard deviation).
SegM1: standard nnU-Net; SegM2: modified nnU-Net with Hausdorff distance loss.
*P<0.05 in paired t-test or Wilcoxon signed-rank test comparing with our method.

**Table 3.** Segmentation performance on real non-contrast CT scans (n=36). All Pearson correlations are significant with p<0.001.

| Method | CUT(DCE)+SegM1 | | | Ours (CUT(DCE)+SegM2) | | | CycleGAN+SegM2 | | | CUT(NCE)+SegM2 | | |
|---|---|---|---|---|---|---|---|---|---|---|---|---|
| Cardiac Chamber | MAPE (%) | MPE (%) | r | MAPE (%) | MPE (%) | r | MAPE (%) | MPE (%) | r | MAPE (%) | MPE (%) | r |
| Left Atrium | 10.07 | 6.52 | 0.93 | 9.22 | 4.67 | 0.93 | 8.79 | 3.81 | 0.94 | 14.80 | 1.08 | 0.82 |
| Left Ventricle | 34.75 | 29.37 | 0.73 | 20.79 | -2.17 | 0.82 | 18.60 | 1.21 | 0.82 | 24.33 | 10.80 | 0.76 |
| Right Atrium | 12.38 | 2.60 | 0.90 | 12.60 | -1.10 | 0.87 | 12.79 | -1.98 | 0.86 | 18.00 | -5.43 | 0.77 |

| Right Ventricle | 15.56 | -7.23 | 0.84 | 16.28 | -12.52 | 0.89 | 16.52 | -12.49 | 0.88 | 25.56 | -13.23 | 0.64 |
|---|---|---|---|---|---|---|---|---|---|---|---|---|

**Note:** SegM1: standard nnU-Net; SegM2: modified nnU-Net with Hausdorff distance loss. MAPE: mean absolute percentage error; MPE: mean percentage error; r: Pearson correlation coefficient.

Scatterplots in Fig. 5 illustrate pairwise associations of volume metrics for the four heart chambers, comparing segmentation results from non-contrast CT with the ground truth measurements from contrast-enhanced CT of the same patient. Strong correlations were evident across all four heart chambers. For the LA, most data points were positioned above the x=y line (red), indicating a tendency of underestimation. Conversely, for the RV, most data points fell below the red line, indicating a tendency of overestimation. Data points for the LV and the RA segmented by the modified nnU-Net (illustrated in green) were distributed more symmetrically along the x=y line. However, the nnU-Net model (illustrated in blue) tended to underestimate the left ventricle volume, with most data points scattered above the x=y line.

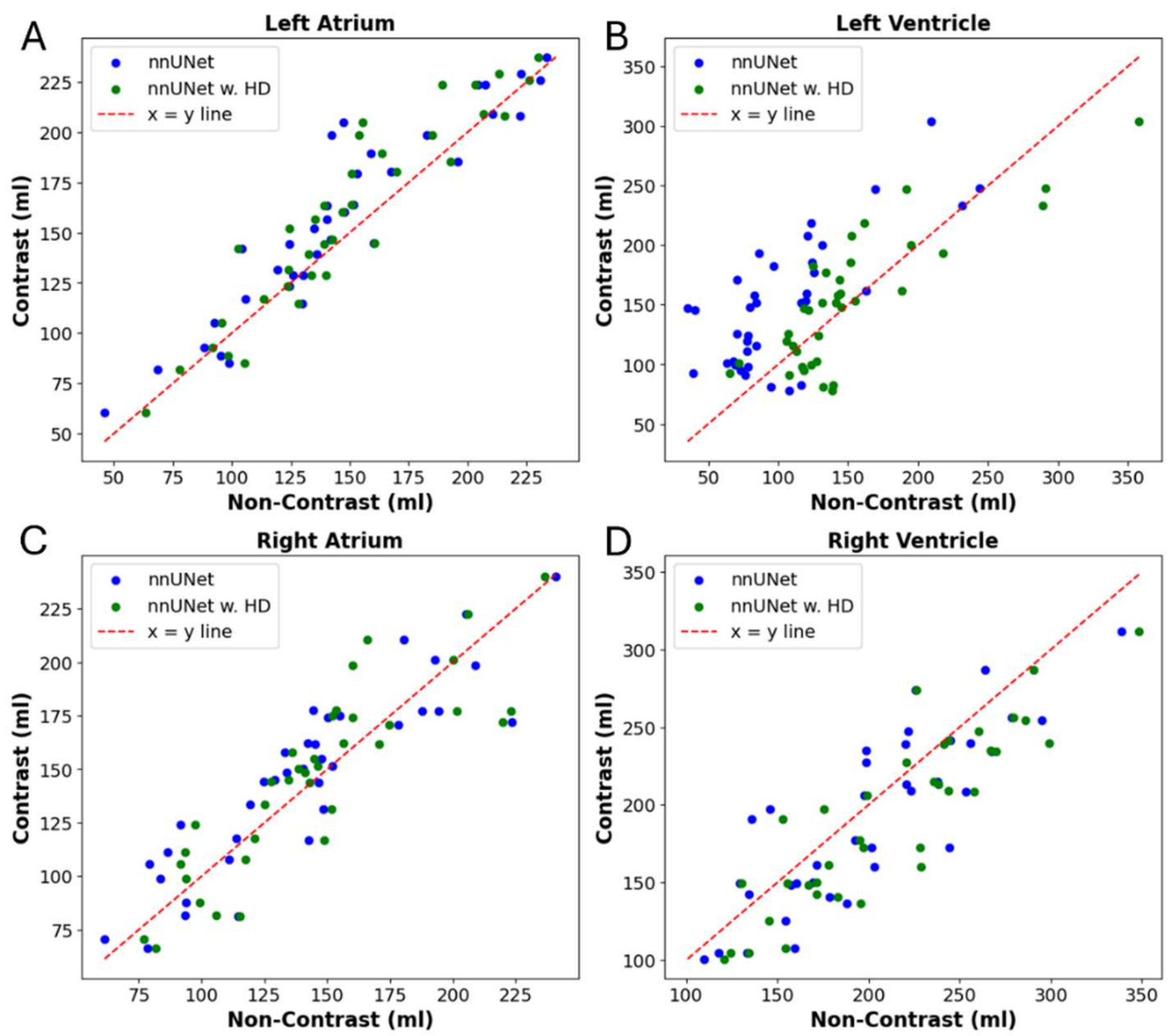

**Figure 5.** Pairwise associations between volumes segmented from real non-contrast CT and ground truth volumes from contrast-enhanced CT using standard nnU-Net and the modified nnU-Net model in the independent test set (n=36). The Pearson correlation coefficient was used to quantitatively evaluate the correlations for volumes of the four heart chambers.

The synthetic and real CT results suggest that Hausdorff distance loss provided limited additional benefit on synthetic images but improved volume agreement when applied to real non-contrast CT scans. Based on these findings, the modified nnU-Net was selected for subsequent comparisons.

### 3.2.2 Effect of Image Translation Method

On synthetic non-contrast CT scans (Table 2), segmentation performance was broadly similar across translation methods. Although CycleGAN+SegM2 and CUT(NCE)+SegM2 showed statistically significant differences in DSC and HD95 compared with our method, the absolute differences were small. On real non-contrast CT scans (Table 3), our method and CycleGAN+SegM2 demonstrated comparable and strong performance across all four chambers. For LA, both methods showed strong correlations with contrast CT-derived volumes (r = 0.93 vs. 0.94), with slightly higher MAPE for our method (9.22% vs. 8.79%). For LV, both achieved r = 0.82, with MAPE of 20.79% for our method and 18.60% for CycleGAN+SegM2. For RA and RV, our method showed slightly higher correlations and lower MAPE than CycleGAN+SegM2. In contrast, the original CUT method performed worse across all chambers, with lower correlations and higher MAPE values, particularly for RV (r = 0.64, MAPE = 25.56%).

### 3.3 Qualitative visualization

Figure 6 shows two examples of heart chamber segmentation on real non-contrast CT scans using the modified nnU-Net model, with corresponding contrast CT annotations provided for visual comparison.

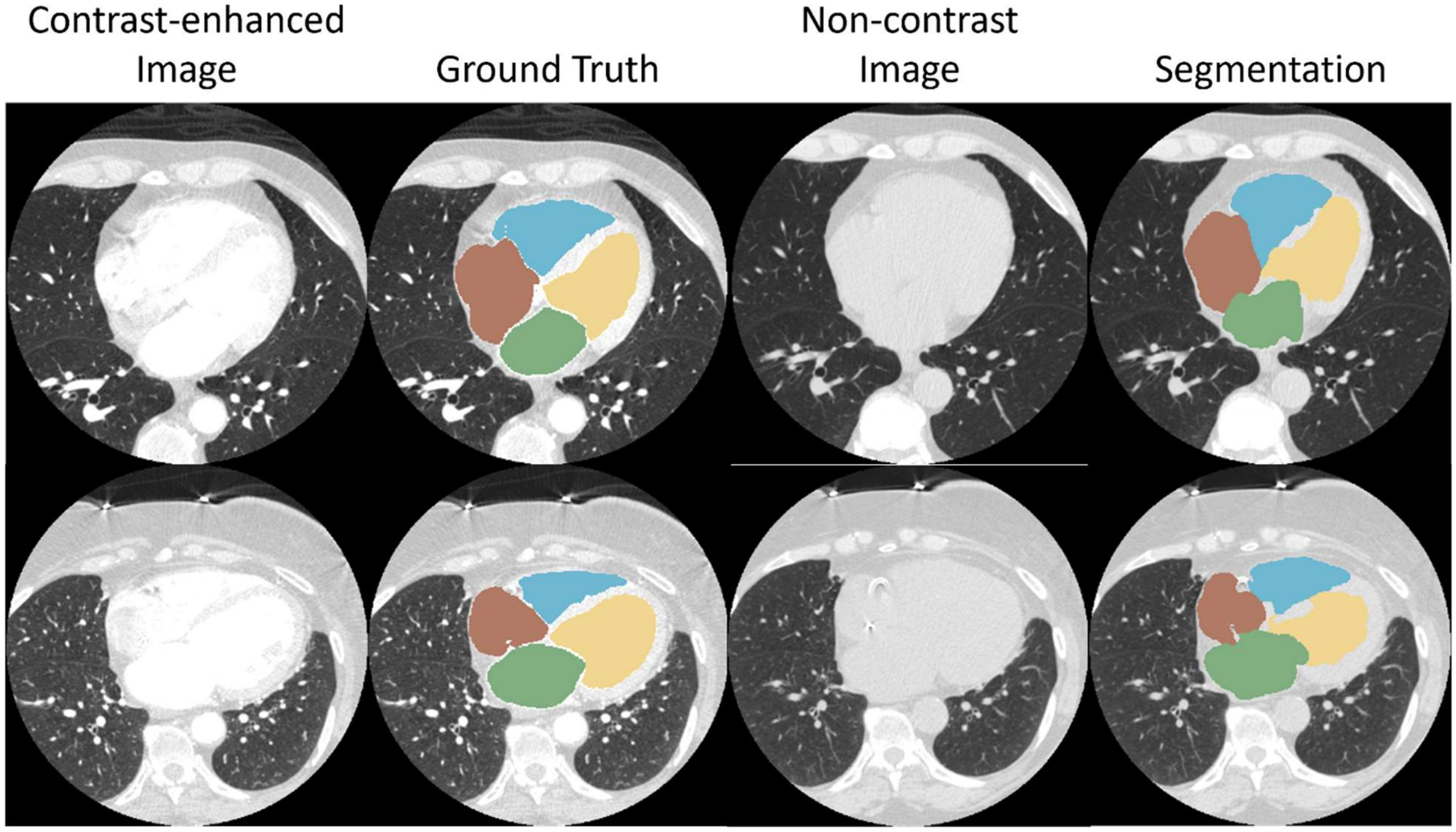

**Figure 6**. Two examples of segmentation results on real non-contrast CT scans by the modified nnU-Net model. Contrast CT images and their corresponding manual delineations on the contrast CT are provided as reference. Columns from left to right: contrast-enhanced images, ground truth on the contrast CT scans, non-contrast CT scans, and the segmentation results on non-contrast CT scans.

## 4. DISCUSSION

In this feasibility study, we developed ChameleonNet, a two-stage framework that combines contrastive unpaired image translation with deep-learning-based segmentation to explore heart chamber segmentation from non-contrast CT scans. The central strategy was to synthesize non-contrast CT images from contrast-enhanced CT scans while preserving anatomical correspondence with existing chamber annotations. This allowed reliable labels from contrast-enhanced CT to be used for training a segmentation model for non-contrast CT, avoiding the need for manual chamber delineation on non-contrast scans, where poor soft-tissue contrast makes annotation challenging.

The image translation stage requires the preservation of anatomical structural integrity, as any distortion could compromise diagnostic reliability and undermine the practical utility of the translated images. While similarity metrics (e.g., pixel-wise error) can help maintain structure integrity in paired translation tasks[28], they present challenges in unpaired translation tasks. This is particularly common in medical settings where paired datasets are often scarce. In cardiac image pairing, factors like patient positioning, breathing, and cardiac motion,  can cause misalignment between two CT scans if techniques such as ECG-gating[33] are not applied. CycleGAN[16] addresses this by using an inverse mapping network and a cycle consistency loss to enforce content consistency. CUT[19] achieves content consistency by maximizing the mutual information between input images and generated images without requiring an additional generator and discriminator. In contrast, our method, following CUT's one-sided design, is more time- and resource-efficient compared to cycle structure (Table 1). In particular, our decoupled contrastive loss further accelerates the training process by eliminating the NPC effect. This allows the generator to quickly learn to produce realistic images that match the target image distribution, which in turn facilitated the tuning of the discriminator.

Due to the unsupervised nature of the translation network used in the first stage, non-contrast and contrast CT scans were randomly selected without enforced alignment. Our method outperformed both CycleGAN (SSIM: 0.93(±0.03), FID: 14.22) and original CUT method (SSIM: 0.95(±0.03), FID: 16.21), with the highest SSIM (0.96±0.03) and lowest FID (13.29) values (Table 1). Although the intensity difference before and after contrast to non-contrast image translation across the three methods was subtle in the central heart area, notable changes along the lung boundaries and within the lungs suggest the superior performance of our method, reflecting the network's learning of global non-contrast CT intensity characteristics (Fig. 4).

The segmentation model integrated a Hausdorff distance loss to regularize segmentation surfaces beyond the standard cross-entropy loss and DSC loss. Model performance was similar with or without the Hausdorff distance loss when testing on the synthesized non-contrast images, yet the Hausdorff distance loss yielded meaningful improvements on real non-contrast CT scans. These gains were most pronounced for the LV, where correlation increased from 0.73 to 0.82, and MAPE was reduced by nearly 14%. This suggests that boundary regularization by the Hausdorff distance loss improves robustness to the difference in intensity distribution between synthetic and real non-contrast CT, which may persist despite low FID. This effect is most notable for chambers like LV, where boundary delineation depends on subtle soft tissue contrast. Comparison across translation methods showed that our method and CycleGAN+SegM2 deliver similar segmentation accuracy on real non-contrast CT across all four chambers despite CycleGAN achieving a higher FID score than our method (14.22 vs. 13.29). This suggests that the FID difference at this scale may not translate to meaningful differences in segmentation-relevant image features, and that segmentation accuracy is not solely determined by translation quality as measured by FID. The original CUT, however, performed substantially worse, consistent with its markedly higher FID score (16.21). Importantly, although our method and CycleGAN+SegM2 achieved similar segmentation results, ours provided superior translation quality and approximately 39% faster training speed, making ChameleonNet the more practical choice for clinical use.

Chen et al.[11] reported DSC values of 0.91(±0.02), 0.93(±0.01), 0.90(±0.02), and 0.91(±0.03), and HD95 values of 3.4 (±0.8) mm, 4.1 (±2.3) mm, 4.4 (±1.5) mm, and 5.6 (±4.5) mm for LA, LV, RA and RV for nnU-Net model on 20 non-contrast CT scans. Bruns et al.[14] reported DSC values of 0.92 (±0.02), 0.90 (±0.03), 0.91 (±0.04) and 0.92 (±0.03) for LA, LV, RA and RV, respectively on VNC images. Our method achieved similar results on the synthesized non-contrast data (Table 2). While direct comparisons between synthesized non-contrast images, real non-contrast images, and the VNC images are not possible, our results demonstrated the segmentation model's ability to learn intrinsic structural information from images with feature distributions similar to real non-contrast images.

Validation on real non-contrast CT scans showed strong agreement with contrast CT-derived chamber volumes from the same subjects, with Pearson correlations of 0.93 for LA, 0.82 for LV, 0.87 for RA, and 0.89 for RV, all $p < 0.001$ (Table 3). These results support the feasibility of applying the trained model to real non-contrast CT scans. The LV showed the greatest improvement with Hausdorff distance loss, suggesting that boundary regularization is particularly helpful for chambers with limited soft-tissue contrast. However, residual errors remained. LV had the highest MAPE among all chambers (20.79%), and RV showed systematic underestimation (MPE = −12.52%). These errors may partly reflect acquisition variability, as contrast and non-contrast scans were not acquired simultaneously and may capture different phases of the cardiac cycle. Consistent with this concern, Carlsson et al.[34] reported an 8.2 ± 0.8% change in total heart volume across the cardiac cycle.

This study has several strengths. First, the model was developed using a relatively large dataset, including 292 contrast CT scans for segmentation training and more than 35,000 axial slices for image translation. Second, the image translation framework does not require paired contrast and non-contrast CT scans, improving its applicability to clinical datasets where paired acquisitions are uncommon. Third, performance was evaluated on an independent held-out set of 36 real non-contrast CT scans, providing a direct feasibility assessment beyond synthesized images. Fourth, the CT-to-CT translation design preserves native CT resolution and intensity information while avoiding the larger modality gap associated with MRI-to-CT adaptation[35]. Finally, the 2D slice-based translation approach reduces memory requirements and supports efficient training, while the 3D segmentation model in the second stage restores volumetric context for chamber delineation.

This feasibility study has several limitations. First, image translation quality was assessed using SSIM and FID, both of which have limitations. SSIM is sensitive to intensity changes, which are expected when translating contrast-enhanced CT to non-contrast CT, making it an imperfect measure of anatomical preservation. FID better reflects feature-distribution alignment between synthesized and real non-contrast CT images, but its use in medical imaging remains debated because it relies on features from an ImageNet-pretrained network[36]. Second, ECG-gated paired contrast and non-contrast CT scans were not available for testing. The contrast and non-contrast scans were acquired from the same subjects but not simultaneously, with intervals ranging from the same day to several weeks. During this time, cardiac volumes may have changed because of physiologic variation, disease status, or treatment. In addition, the absence of ECG gating means that scans may have captured different cardiac phases, introducing unavoidable variability in chamber-volume comparisons and preventing direct Dice-based evaluation on real non-contrast CT scans. Finally, image registration between paired contrast and non-contrast CT scans with elastix[37] was explored but was not reliable. Differences in cardiac phase, physiologic changes between acquisitions, and limited soft-tissue contrast on non-contrast CT introduced spatial inaccuracies and made cardiac boundary alignment difficult. Future studies using ECG-gated paired contrast and non-contrast CT scans are needed to enable direct voxel-level evaluation, more reliable volume comparison, and further refinement of the method before clinical or large-scale research use.

## 5. CONCLUSION

We proposed and validated a novel framework to segment heart chambers from non-contrast CT scans. By synthesizing non-contrast CT scans from contrast CT scans, we were able to transfer annotations from contrast CT to non-contrast CT without the need for paired data. This approach capitalizes on the availability of easily obtained annotations from contrast CT to train a segmentation model on synthesized non-contrast CT images. The validation results on real non-contrast CT scans demonstrate the feasibility of estimating cardiac chamber volumes on non-contrast CT scans. However, chamber-specific volume errors, particularly for the LV and RV, indicate that performance on real non-contrast CT remains imperfect, highlighting the need for further refinement and validation before clinical or large-scale research use.

**Acknowledgement:**

This work is supported in part by research grants from the National Institutes of Health (NIH) (R01CA237277 and R01HL174570).

**Disclosures:**

The authors declare that there are no financial interests, commercial affiliations, or other potential conflicts of interest that could have influenced the objectivity of this research or the writing of this paper.

**Code and Data Availability:**

The code of this study can be accessed through Github repositories:

jingW-0/contrast2noncontrast

jingW-0/nnUNet_customize

The data cannot be shared publicly as they consist of identifiable patient information obtained from institutional hospital databases.